\documentclass[eqsecnum,preprint,showpacs,aps]{revtex4}
\usepackage{epsfig}
\def\beq{\begin{equation}}
\def\eeq{\end{equation}}
\def\bea{\begin{eqnarray}}
\def\eea{\end{eqnarray}}
\def\nnu{\nonumber}
\def\lf{\left}
\def\rt{\right}
\def\tst{\textstyle}

\def\fno#1{Fig.~\ref{#1}}

\def\eno#1{Eq.~(\ref{#1})}
\def\Eno#1{Equation (\ref{#1})}
\def\etwo#1#2{Eqs.~(\ref{#1}) and (\ref{#2})}

\def\gam{\gamma}
\def\dta{\delta}
\def\eps{\epsilon}
\def\tta{\theta}

\def\lam{\lambda}

\def\Gam{\Gamma}
\def\Dta{\Delta}

\def\by{\over}

\def\apx{\approx}
\def\ptl{\partial}

\def\hf{{1\over2}}
\def\tshf{\tst\hf}

\def\ham{{\cal H}}
\def\ket#1{|#1\rangle}

\def\tran#1#2{\langle#1|#2\rangle}
\def\avg#1{\langle#1\rangle}
\def\mel#1#2#3{\langle#1|#2|#3\rangle}

\def\bH{{\bf H}}
\def\bJ{{\bf J}}

\def\xhat{{\bf{\hat x}}}
\def\yhat{{\bf{\hat y}}}
\def\zhat{{\bf{\hat z}}}
\def\nhat{{\bf{\hat n}}}

\def\Fe8{Fe$_8$}
\def\Mn12{Mn$_{12}$}

\def\baz{{\bar z}}
\def\bat{{\bar\tta}}
\def\bap{{\bar\phi}}
\def\bzf{{\bar z}_f}
\def\Scl{S^{{\rm cl}}}
\def\emin{E_{\rm min}}

\def\bl2{{\bar\lam}_2}

\begin{document}

%\twocolumn[
%\hsize\textwidth\columnwidth\hsize\csname@twocolumnfalse\endcsname

\title{Spin Tunneling in Magnetic Molecules: Quasisingular
Perturbations and Discontinuous SU(2) Instantons}

\author{Ersin Ke\c{c}ecio\u{g}lu}
\author{Anupam Garg}
\email[e-mail address: ]{agarg@northwestern.edu}
\affiliation{Department of Physics and Astronomy, Northwestern University,
Evanston, Illinois 60208}

\date{\today}

\begin{abstract}
Spin coherent state path integrals with discontinuous semiclassical
paths are investigated with special reference to a realistic model for
the magnetic degrees of freedom in the \Fe8 molecular solid. It is
shown that such paths are essential to a proper understanding of the
phenomenon of quenched spin tunneling in these molecules. In the \Fe8
problem, such paths are shown to arise as soon as a fourth order
anisotropy term in the energy is turned on, making this term a singular
perturbation from the semiclassical point of view. The instanton
approximation is shown to quantitatively explain the magnetic field
dependence of the tunnel splitting, as well as agree with general rules
for the number of quenching points allowed for a given value of spin.
An accurate approximate formula for the spacing between quenching points
is derived.
\end{abstract}

\maketitle

%\widetext
\section{Introduction}
\label{intro}
\subsection{Overview}
\label{oview}
Among the several advances in the study of large-spin molecular magnets
in the last few years~\cite{vill}, perhaps the most remarkable is the
observation of quenching of spin tunneling in the molecule
[Fe$_8$O$_2$(OH)$_2$(tacn)$_6$]$^{8+}$ (or just \Fe8 for brevity) by Wernsdorfer
and Sessoli (WS)~\cite{werns,mn12new}. The spin of this molecule can tunnel
between classically degenerate orientations, but for certain values of the
applied magnetic field, it is found that the tunneling is eliminated, or
quenched. Between quenching points, the tunnel splitting oscillates as a
function of applied field. The simplest model that is capable of describing
this effect has been studied by many workers now~\cite{gargepl,gargprl99,%
kou,fort,lmpp,yl,gargmathph,gargprb01}, and it is well understood qualitatively
in terms of an interference between tunneling Feynman paths. Such interference
can also arise for massive particles in more than one dimension~\cite{wilk}. It arises
in the case of a single spin because the kinetic term in the path integral has
a geometrical or Berry phase structure~\cite{ldg,vdh,gargepl}.

There is, however, yet another aspect of the \Fe8 experiments that has not been widely
appreciated, because of which the physical explanation for the tunnel
splitting oscillations is significantly incomplete. This is because if we view
the tunneling in terms of just the interfering
paths, the geometrical interpretation of the relative phase between them
inevitably predicts a certain number --- ten --- of quenching points. Experimentally,
and by numerical diagonalization of the appropriate model Hamiltonian, only four
quenching points are seen.
We will show in this paper
that the resolution of this apparent paradox requires that we include paths that
have {\it discontinuities\/} at the end points. As a general point, the
necessity of including such paths in the spin-coherent-state path integral
has in fact been known for some time~\cite{Fadd,Klau}, but the significance
of this idea has not progressed much beyond the purely formal level, and the one
concrete problem where discontinuous paths are known to be needed --- Larmor precession in a
constant magnetic field --- is so well understood to begin with, that few people have
pursued this line of inquiry further. The \Fe8 problem has provided us with a strong
motivation for doing just that, thus enhancing our understanding of
the spin-coherent-state path integral. In turn, we can now see
that discontinuous paths exist in all other coherent-state path integrals
too, and this may prove to be of use in situations where such path integrals are the
natural analytic tool.

There are, therefore, two points to this paper. First, as already stated, is to
advance our understanding of the spin-coherent-state path integral. Second, we obtain
a {\it quantitative\/} analytical explanation for additional aspects of WS's data
that are not captured by the simple model, and that have up till
now been only understood numerically. This is desirable as the oscillations seen in
\Fe8 are, to date, the clearest evidence for spin tunneling
of a spin of such large magnitude.

This may be a good point to ask why one should bother with an analytical
explanation at all.
After all, the model Hamiltonian (\ref{ham1}) given below can be numerically
diagonalized with minimal effort. A numerical diagonalization by itself gives no
insight into the results, however, and if it were the only way we had of
solving the problem, the oscillations in the splitting
would be a complete mystery. If one believes that the mental picture of interfering
tunneling paths is useful, then it is surely important to know if and when and how
this picture breaks down. Secondly, the numerical approach cannot by itself explain
the scale of the tunnel splitting.

A shorter paper with our results has been published previously~\cite{ek+ag2}.
In the present paper, in addition to providing the details of the work,
we explain how to find the end points of the discontinuous paths or
boundary jump instantons, as we refer to them. We also discuss the general
model for parameter values other than those relevant to \Fe8 in order to
further test the scenario for the elimination of quenching points. Finally, we
show that the number of quenching points that remain must be even or odd depending
on the value of the spin, $J$, and how this
comes about from the instanton analysis.

\subsection{Basic Facts about \Fe8}
\label{fe8facts}
The main facts about \Fe8 are as follows. The cluster of
eight Fe$^{3+}$ ions in one molecule has an approximate $D_2$ symmetry. 
Competing exchange interactions among the ionic spins in one molecule lead to
a ground state with a total spin of $J=10$. The compound forms a well ordered
crystalline solid, in which the magnetic ions in one molecule are
kept well separated from those in another molecule by large organic
ligands. The magnetic degrees of freedom may thus be treated as a set of
non-interacting spins, each with $J=10$. To a first approximation, the
anisotropy of each molecule may be described by the Hamiltonian
\beq
\ham_0 = - k_2 {J_z}^2 + (k_1- k_2) {J_x}^2 - g \mu _B \bJ\cdot\bH.
\label{ham} 
\eeq
Here, $g$ is the gyromagnetic ratio, $\mu_B$ is the Bohr magneton, and 
$k_1 > k_2 > 0$ are anisotropy parameters. \Eno{ham} is the simplest
Hamiltonian that can describe the quenching effect.
Viewed as a classical Hamiltonian, it has two degenerate ground states at the
points $\bJ = \pm J\zhat$ in the absence of an applied field. These states are
separated by a barrier $k_2J^2$ along the $y$-axis, and $k_1J^2$ along the
$x$-axis. Quantum mechanics admixes these states via tunneling~\cite{zero_pt}. 
When the field is non-zero, the classical minima are no longer along $\pm\zhat$,
but they are still degenerate if $\bH \perp\zhat$, the easy axis, and one can
have tunneling between the corresponding quantum mechanical states. More generally,
one can have tunneling between excited states in the two wells. In this paper,
however, we will only consider $\bH\| \xhat$, and tunneling between the lowest
two states. In this case, the quenching points are perfectly regularly spaced,
and are located at 
\beq
H_x = (J - n - \tshf) \Dta H_x^{(0)}, \quad n = 0, 1, \ldots, 2J - 1,
   \label{DPloc}
\eeq
where the spacing or period is given by
\beq
\Dta H_x^{(0)} = {\sqrt{1-\lam} \over J}H_c, \label{per0}
\eeq
where $\lam = k_2/k_1$, and $H_c = 2 k_1 J/g\mu_B$. (These, and other important
parameter combinations are tabulated in Table \ref{parcomb}.)
\begin{table}[t]
\caption{\label{parcomb}Important parameter combinations}
\begin{ruledtabular}
%\vspace{0.2cm}
%\begin{center}
\begin{tabular}{c c}
Quantity & Formula \\
\hline
$H_c$                    &    $2k_1 J/g\mu_B$     \\
%$u_0$, $\cos\tta_0$      &    $H/H_c$             \\
$\lam$                   &    $k_2/k_1$           \\
$\lam_2$                 &  $CJ^2/k_1$            \\
$h$                      &  $H/H_c$               \\
$\zeta$                  &  $4\lam_2 h^2$         \\
\end{tabular}
\end{ruledtabular}
\end{table}

We note in passing that although this and similar results for other quenching
points were initially obtained
by various semiclassical approximations \cite{gargepl,gargprl99,fort},
they are in fact exact~\cite{ersin} for the model (\ref{ham}).

WS's observations differ from the simple model predictions in that they see
only four quenching points on the positive $H_x$ axis, where the simple model
would say ten, and the spacing between these points is more than 50\% greater
than the formula (\ref{per0}) would give. (See \fno{splitting}.)
\begin{figure}
\epsfig{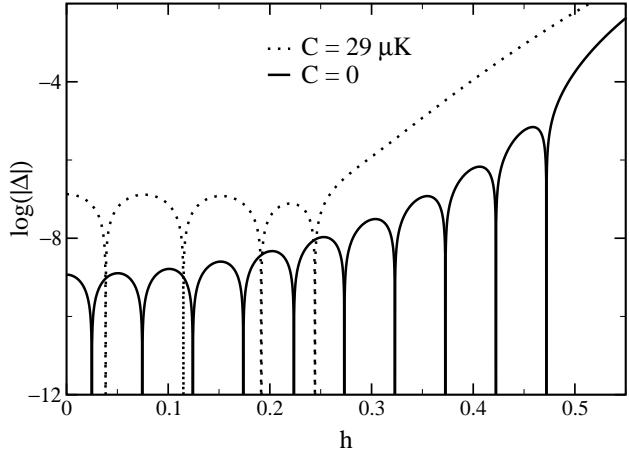}
\caption{\label{splitting} 
Tunnel splitting $\Dta$ between the ground level pair for the Hamiltonian
(\ref{ham1}) with $C=0$ (solid line) and $C = 29\,\mu$K (dotted line).
Plotted is $\log_{10} |\Dta|$, with $\Dta$ measured in Kelvin. 
                                   }
\end{figure}
At the same time, this spacing is quite
uniform. They surmise that the differences are due to the presence of higher
order anisotropy terms in the Hamiltonian consistent with the symmetry of the
molecule, and by trial and error and numerical diagonalization of a
$21 \times 21$ matrix, they discover that their data can be very well fit by
the following model Hamiltonian \cite{expl}
\beq
 \ham = -k_1 J_z^2 + (k_1-k_2) J_x^2 - C [J_{+}^4 + J_{-}^4]
         - g \mu _B \bJ\cdot\bH.
 \label{ham1}
\eeq 
The parameters $g$, $k_1$ and $k_2$ are known through a variety of
experimental evidence~\cite{gatt,barra,cac,henn}. We will use the values
used by WS: $g\simeq 2$, $k_1 \simeq 0.338$~K, $k_2\simeq 0.246$~K, and
$C = 29\ \mu$K. It turns out to be important that one needs $C>0$.

This, then, describes the problem that we wish to solve. One point to note is
that the dimensionless strength of the fourth order anisotropy
$C$ is
\beq
\lam_2=CJ^2/k_1.
\eeq
For the \Fe8 parameter values, $\lam_2 = 8.580 \times 10^{-3}$. That such a
small term should have such a large effect (elimination of six quenching
points, 50\% increase in period), suggests that it is a singular perturbation.
We shall see that this is indeed true when we analyze its effect on the
semiclassical paths, although the perturbation is perfectly well behaved
from the quantum mechanical operator point of view.

\subsection{Plan of paper}
\label{plan}

The plan of the paper is as follows. In Sec.~\ref{instformal} we briefly
review the formalism of SU(2) instantons with special emphasis on the
jump instantons. To this end, we first discuss (Sec.~\ref{scsparam})
stereographic coordinates
for the unit sphere, the corresponding parametrization of spin coherent
states, and how paths on the complexified unit sphere are represented
in terms of stereographic as well as the usual spherical polar coordinates.
We then discuss (Sec.~\ref{spinaction}) the action functional for spin,
and show that unless one
includes a ``boundary term" that depends explicitly on the boundary values of
the spin path, there is no sensibly formulatable least action principle, nor
is there a solution to the Euler-Lagrange equations in general. Paths with
discontinuities arise naturally once a boundary term is included. We also
discuss Klauder's formulation of these points in terms of an extra
infinitesimal pseudo-inertial term in the kinetic energy and the concomitant
production of narrow boundary layers in the spin paths. In the limit where
the extra kinetic energy term vanishes, the width of these layers also
vanishes, and they then look like discontinuities in the paths.

In Sec.~\ref{strucans}, we discuss the general structure of the tunnel
splitting for the model (\ref{ham1}) when both types of instantons---with
and without jumps---are allowed. We shall see that an instanton of the
latter type dominates as the field gets large, and since this instanton
has no interfering partner, we see why the splitting ceases to oscillate
after fewer than ten quenching points have been realized.

In Sec.~\ref{fe8anal} we present the explicit analysis for our
model for \Fe8, \eno{ham1}. In Sec.~\ref{period} we analyse the interfering
instantons. We identify $\zeta = 4\lam_2 h^2$ (with $h = H/H_c$)
as the appropriate small
parameter, and explain why the $H_x$ spacing of the quenching points is so
regular. We obtain a concrete formula for this spacing, namely,
\beq
\Dta H_x = {\pi H_c \over J I(\lam,\lam_2)}, \label{per1}
\eeq
where $I$ is an integral expression [see \eno{perint}] depending
on the parameters $\lam=k_2/k_1$, $\lam_2=CJ^2/k_1$ (see also
Table \ref{parcomb}).
With the parameter values relevant to \Fe8, we obtain $I=3.88$, giving
$\Dta H_x = 0.409$~T. WS quote a period of 0.41~T. Compared to the model
$\ham_0$, the period is increased by a factor $\pi (1-\lam)^{-1/2}/I = 1.552$.
We relate $I$ to a complete elliptic integral, and give various formulas
for it in the Appendix.
%\ref{intappend}. 

In Sec.~\ref{bjinsts} we turn to the jump instantons. We first formulate
the equations for determining the endpoints of these instantons. In general
these equations would have to be solved numerically, but for the \Fe8
model, it turns out that an analytical solution can be found. With
these endpoints in hand, we can find the complete instantons and their
actions (this must still be done numerically). We use our computed
actions for all the instantons to find the splitting as a function of
magnetic field, and compare with the numerical diagonalization. This
exercise is carried out for the value of $\lam_2$ applicable to \Fe8,
as well as some others.

In Sec.~\ref{Nquench} we address the issue of the number of quenching points
in light of general quantum mechanical theorems. It is not an accident that
WS see an even number (four) of quenchings for $H_x > 0$. We will see that
the general theorems constrain this number for any value of $J$, and then
show how this same conclusion comes about from the instanton analysis.

We conclude the paper in Sec.~\ref{concl} with a summary and discussion.

\section{SU(2) Instanton Formalism Revisited}
\label{instformal}
The instanton method is an efficient way of calculating tunnel splittings,
both for particles~\cite{Fel} and for spin~\cite{hanbook}. It is based on
evaluating the path integral for a certain propagator in the steepest-descent
approximation, and is designed to be asymptotically correct in the semiclassical
limit ($J\to\infty$ or $\hbar \to 0$). Instantons are classical paths that run
between degenerate classical minima of the energy. By ``classical" we mean that
the path obeys the principle of least action, and, {\it a fortiori\/}, satisfies
energy conservation. However, a path along which energy is conserved can not run
between two minima and still have real coordinates and momenta. Hence, one must
enlarge the notion of a classical path, and allow the coordinates and/or momenta
to become complex. One is naturally led in this way to reexamine the least action
principle. In the case of spin, there is yet another reason for this
reexamination, as we shall see.

\subsection{Spin coherent states and their parametrization}
\label{scsparam}
More specifically, let the degenerate classical energy minima be along the
directions $\nhat_i$ and $\nhat_f$. We seek the propagator
\beq
K_{fi} = \mel{\nhat_f}{\exp[-\ham T]}{\nhat_i} \label{Kfimel}
\eeq
in the limit $T \to \infty$.  Here, $\ket{\nhat_{i,f}}$ are spin coherent states
(defined below). This propagator is given by the path integral
\beq
K_{fi} = \int {\cal D}[\nhat(t)] e^{-S[\nhat(t)]}, \label{Kfipath}
\eeq
where $\nhat(0) = \nhat_i$, $\nhat(T) = \nhat_f$, and $S$ is the (Euclidean)
action. When we evaluate this integral by steepest
descents, least action paths emerge naturally.

We define the spin coherent state $\ket\nhat$ as the state with maximal spin
projection along the direction $\nhat$. In other words, $\nhat$ is an
eigenstate of $\bJ\cdot\nhat$ with eigenvalue $J$:
\beq
\bJ\cdot\nhat \ket\nhat = J \ket\nhat. \label{def_scs}
\eeq
The most common way of explicitly writing $\ket\nhat$ and $S[\nhat(\tau)]$ is in terms
of the spherical polar coordinates $\tta$ and $\phi$ of the direction $\nhat$. For
formal purposes, another representation is more convenient. Let $z$ be a complex
number, related to $(\tta,\phi)$ by the stereographic map
\beq
z = \tan{\tta \over 2} e^{i\phi}. \label{defstereo}
\eeq
Then, up to normalization and phase, $\ket\nhat$ is identical to the state~\cite{fnJJ} 
\beq
\ket{z} = e^{zJ_-} \ket{J,J}. \label{defketz}
\eeq
The advantage of this representation is that matrix elements of various operators
have nice analyticity properties. For example, for Hamiltonians polynomial
in the components of $\bJ$,
\beq
H({\bar z}', z) = {\mel{z'}{\ham}{z} \over \tran{z'}{z}}
\eeq
is holomorphic in $z$ and antiholomorphic in $z'$, where $\baz$ is the formal
complex conjugate of $z$. In equations, this means that
\beq
{\ptl \over \ptl {\bar z}} H({\bar z}', z) = 
{\ptl \over \ptl z'} H({\bar z}', z) = 0. \label{Hanal}
\eeq
Explicit model calculations, however, are often easier in $\tta$ and $\phi$.

Since we will be considering paths on the complexified unit sphere, it is
useful to understand what this means in terms of the two coordinate systems.
Let $u$, $v$, and $w$ be Cartesian coordinates in three-dimensional space.
The real unit sphere is the surface specified by
\beq
u^2 + v^2 + w^2 =1. \label{unitsph}
\eeq
The complexified unit sphere is obtained by allowing $u$, $v$, and $w$ to
become complex. The real and imaginary parts of \eno{unitsph} provide two
constraints among six variables (the real and imaginary parts of $u$, $v$,
and $w$), leaving us with a four-dimensional manifold. We can also see this
in the terms of polar and stereographic variables. Consider the former
first. We relate them to Cartesian coordinates in the usual way
\beq
(u, v, w) = (\cos\tta, \sin\tta\cos\phi, \sin\tta\sin\phi).
\eeq
\Eno{unitsph} is automatically satisfied. This continues to be true if we
allow $\tta$ and $\phi$ to be complex. Once again, we conclude that four
real quantities are required to specify a point on this manifold. Let us
consider stereographic variables next. Since
\beq
z = \tan{\tta\by 2}e^{i\phi}, \quad 
\bar z = \tan{\tta\by 2}e^{-i\phi},
\eeq
we see that if $\tta$ and $\phi$ are complex, $\bar z$ will not be the
same as $z^*$, the true complex conjugate of $z$. To specify a point on
the complex unit sphere, both
$z$ and $\bar z$ are needed, i.e., four real parameters are needed. Conversely,
a point with stereographic coordinates $(\bar z, z)$ lies on the real unit
sphere (which is a submanifold of the complex unit
sphere) if $z$ and $\bar z$ are complex conjugates. Such points may be given by
specifying $z$ alone or $\bar z$ alone. Thus, corresponding to the directions
$\nhat_i$ and $\nhat_f$, which are real, we may speak of ``the points" $z_i$
and $z_f$, or ``the points" ${\bar z}_i$ and ${\bar z}_f$.

\subsection{Action functional for spin}
\label{spinaction}

Let us now consider the action.
To specify a path $\nhat(t)$ on the unit sphere stereographically,
one must give both $z(t)$ and $\baz(t)$. The action (or more precisely,
the Hamilton principal function) for a path obeying the boundary conditions
\beq
z(0) = z_i, \quad \baz(T) = \bzf, \label{bczz}
\eeq
is given by~\cite{koch,spg}
\beq
S \lf( \baz(t), z(t) \rt ) = S_K + S_D + S_B, \label{skbd}
\eeq
where
\bea
S_K &=& -\int_0^T \lf[ J {\dot{\baz} z - \baz{\dot z} \over 1 + \baz z}
                           \rt] dt \label{Skin}\\
S_D &=& \int_0^T H(\baz, z) dt \label{Sdyn} \\
S_B &=& - J \ln \lf[{(1+ \baz(0) z_i) (1 + \bzf z(T))
                               \over
                            (1+ z^*_i z_i) (1 + z^*_f z_f)}\rt].
      \label{Sbndry}
\eea
We refer to these terms as the kinetic, dynamical, and boundary terms,
respectively. In $S_B$, $z^*_{i,f}$ denotes the {\it true\/} complex
conjugate of $z_{i,f}$. This term depends explicitly on the boundary values
of the path, and is needed to avoid the overdetermination
problem~\cite{Fadd,Klau}. We state why this is so without proof.
If we vary the path $\baz(t)$, $z(t)$,
{\it including the endpoints\/}, and require the resulting variation
$\dta S$ to vanish, we discover (i) the Euler-Lagrange (EL) equations
\beq
\dot\baz = {(1 +\baz z)^2 \over 2J} {\ptl H \over \ptl z}, \qquad
\dot z = -{(1 +\baz z)^2 \over 2J} {\ptl H \over \ptl\baz}, \label{ELzz}
\eeq
and (ii) that $\dta S$ has no terms proportional to $\dta \baz(0)$ or $\dta z(T)$.
This means that the action evaluated for the classical path $z_{\rm cl}(t)$,
$\baz_{\rm cl}(t)$ [which obeys \etwo{bczz}{ELzz}] is not a function of $\baz_i \equiv \baz(0)$ and
$z_f \equiv z(T)$:
\beq
{\ptl \over \ptl \baz_i} \Scl = 
{\ptl \over \ptl z_f} \Scl = 0. \label{Sanal}
\eeq
Equivalently, one can say that
\beq
S[\baz_{\rm cl}(t), z_{\rm cl}(t)] = \Scl(\baz_f, z_i, T),
\eeq
where we show the variables on which $\Scl$ depends explicitly. Third, one obtains
the Hamilton-Jacobi equations
\beq
{\ptl \Scl \over \ptl \baz_f} = 2J {z(T) \over 1 + \baz_f z(T)}, \quad
{\ptl \Scl \over \ptl z_i} = 2J {\baz(0) \over 1 + \baz(0) z_i}.
\eeq
Lastly, the ``energy" is conserved along the classical path:
\beq
{d \over dt} H\bigl( \baz_{\rm cl}(t), z_{\rm cl}(t)) = 0. \label{econs}
\eeq
This follows from \eno{ELzz}.

What would have happened if we had omitted the boundary term $S_B$ in
\eno{skbd}? Since a general variation $\dta S$ would include terms proportional
to all four quantities $\dta z_i$, $\dta\baz_i$, $\dta z_f$, and $\dta\baz_f$,
the classical path would have to be specified via four boundary conditions
$(z_i, \baz_i, z_f, \baz_f)$, whereas the EL conditions would still
form a system of differential equations of order 2. The system would be
overdetermined with no solution in general.
(The same problem can be seen in spherical polar coordinates. One must give
$\tta_i$ and $\phi_i$ to specify $\nhat_i$, and $\tta_f$ and $\phi_f$
for $\nhat_f$, i.e.,
four boundary conditions in all.) Second, even with only two boundary conditions
$z(0) = z_i$, $\baz(T) = \baz_f$, for general $z_i$ and $\baz_f$ there is
no {\it real\/} solution to the EL equations. That is to say, $\baz_{\rm cl}(t)$
is not equal to the true complex conjugate of $z_{\rm cl}(t)$. A solution will
in general exist only if we allow $\baz_{\rm cl}(t)$ and $z_{\rm cl}(t)$ to be
independent complex functions. [In spherical polar coordinate language, we must
allow both $\tta(t)$ and $\phi(t)$ to become complex.]
We thus see that if we want to consider classical dynamics for spin from the
viewpoint of a least action principle, we must allow the dynamics to
live on the complexified unit sphere from the very outset.

We have couched the above discussion in language following Faddeev~\cite{Fadd}.
It is also useful to discuss Klauder's resolution of these issues~\cite{Klau}.
On the face of it, his treatment is slightly different, but turns out to be
equivalent to Faddeev's in actual applications. Klauder does not include
an explicit boundary term in $S$, but argues that since the
continuum path integral is a formal construct with meaning only as a limit of
its discrete version, one may add a term to the integrand for $S_K$ that is
quadratic in the velocities $\dot\tta$ and $\dot\phi$, and which has an
infinitesimal
coefficient $\eps$. The EL equations are then a fourth order system, and a
classical solution always exists. However, it has the following structure.
It evolves from $(\tta_i,\phi_i)$ to a point $(\bat_i,\bap_i)$ (note the different
use of the overbar) in a boundary layer of duration $O(\eps)$, evolves for a time
$T - O(\eps)$ along a path $\bat(t), \bap(t)$ according to
\beq
iJ \sin\tta {d \tta \over dt} = {\ptl H \over \ptl \phi}, \quad
iJ \sin\tta {d \phi \over dt} = -{\ptl H \over \ptl \tta}, \label{ELtp}
\eeq
and then evolves from a point $(\bat_f, \bap_f)$ to $(\tta_f, \phi_f)$ in
another boundary layer of duration $O(\eps)$ near $t = T$.
[Note that \eno{ELtp} is equivalent to \eno{ELzz}.] The extra kinetic term
in the action gives a non-zero contribution only from integration over the
boundary layers, but this contribution is explicitly independent of $\eps$
as $\eps \to 0$. The net classical action is then given by
\bea
\Scl &=& \int_0^T \lf[ iJ(1 - \cos\bat)\dot\bap + H(\bat,\bap) \rt] dt\nnu\\
     && + 2J \ln \lf[{ \cos\tshf\bat_i \cos\tshf\bat_f
                                \over
                             \cos\tshf\tta_i \cos\tshf\tta_f} \rt]. \\
\label{SEjump}	 
\eea
The last term is the boundary layer contribution, and is equivalent to the
explicit term $S_B$ in \eno{skbd} \cite{fnbndry}.
The boundary values $(\bat_i,\bap_i)$ and $(\bat_f, \bap_f)$ are constrained by
\bea
\tan \tshf\bat_i e^{i\bap_i} &=& \tan \tshf\tta_i e^{i\phi_i}, \label{bczi} \\
\tan \tshf\bat_f e^{-i\bap_f} &=& \tan \tshf\tta_f e^{-i\phi_f}, \label{bczf}
\eea
[note the similarity to $z(0) = z_i$, $\baz(T) = \baz_f$]. Once again
a solution can in general be found only if $\bat(t)$ and $\bap(t)$ are
complexified. 
 
\subsection{Structure of answer for \Fe8}
\label{strucans}

Once the action functional is specified, the instanton recipe
for calculating the tunnel splitting is as follows. Let there be a
number of instantons, i.e., least action paths, labeled by $k$, and let
the actions for these various paths be $\Scl_k$. The tunneling amplitude
is given by
\beq
\Dta = \sum_k D_k e^{-\Scl_k}.
\eeq
This can be understood as arising from the path integral (\ref{Kfipath})
in the following way. In the integral, the dominant paths (in the sense
of the method of steepest descents), are those for which the action is
stationary. These are responsible for the exponential factor
$\exp(-\Scl_k)$. The prefactor $D_k$ results from
integrating out the Gaussian fluctuations around the $k$th instanton. 
The full expression for $K_{fi}$ must include a sum over multi-instanton
paths, and this leads to a result which is essentially exponential in $\Dta$,
in a way which is now well understood \cite{Fel}.

Up to now, all explicit instanton calculations of tunnel splittings of
which we are aware have the property that the initial and final points
of the instantons lie on the real unit sphere. In other words,
$\baz(0) = z^*_i$, $z(T) = \baz^*_f$. In
Klauder's language, $(\bat_{i,f},\bap_{i,f}) = (\tta_{i,f}, \phi_{i,f})$, so
there are no boundary layers in the path.
(At intermediate times, of course, the path {\it is\/} complexified.) This is
due to simplifying special features present in the models studied. As a result,
there is no need to include the explicit boundary terms in $S$ or $\Scl$, and
the practice in all papers on spin instantons (including those written by one
of us, AG), has been to forget about them altogether. In the present
problem, however, we find that although instantons with end points on the
real sphere still exist, one also has instantons for which this is not so.
We refer to these as {\it boundary jump instantons\/} (BJI). It is essential
to include the latter in order to understand why the number of diabolical points
on the $H_x$ axis is reduced from 10 to 4.

More specifically, we have four instantons, labeled 1, 2, 3 and 4
(see \fno{phasepath}).
\begin{figure}
\epsfig{figure=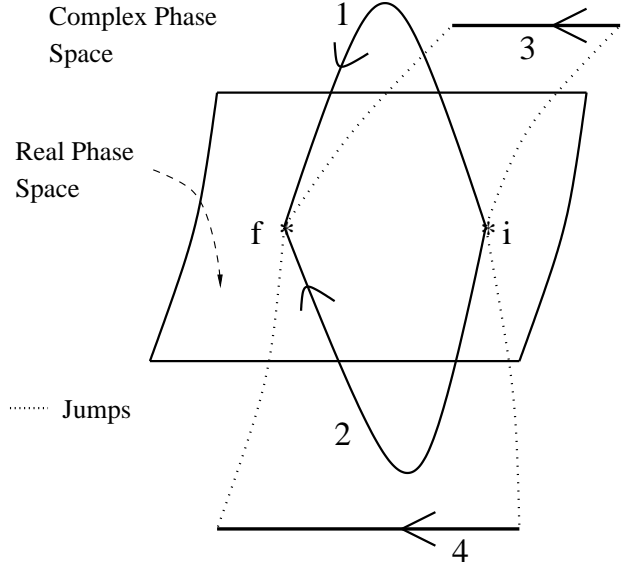,width=0.5\linewidth}
\caption{\label{phasepath} 
Schematic of instanton paths in \Fe8. We show a portion of real phase space
(the real unit sphere) as a two-dimensional surface, that is embedded in 
complex phase space. The latter is four-dimensional, but we can only
represent it as three-dimensional in this perspective drawing. The initial
and final points {\it i\/} and {\it f\/} lie in real phase space. Paths 1 and 2
start at these points, have no jumps, interfere with
each other, and evolve smoothly as the $C$ term in \eno{ham1} is turned on.
Except at the end points these paths lie in complex phase space.
Paths 3 and 4 possess jumps at the end points, do not interfere, and
are obtained only when $C \ne 0$. These paths lie entirely in complex
phase space.}
\end{figure}
The first two have no boundary jumps, and are the ones that interfere as in
Ref.~\cite{gargepl}. The third and fourth are boundary jump instantons.
Hence,
\beq
\Dta = \sum_{k=1}^4 D_k e^{-\Scl_k}.
\eeq
By proper choice of gauge, one can ensure that all $D_i$ are real,
$\Scl_3$ and $\Scl_4$ are real, and that $\Scl_2 = (\Scl_1)^*$, $D_2 = D_1$.
More generally, the real parts of $\Scl_1$ and $\Scl_2$ must be equal to
each other by symmetry, and the imaginary parts must be related by
\beq
\Scl_2 - \Scl_1 = 2 i J \Phi, \label{BP}
\eeq
where $\Phi$ is half the area on the complexified unit sphere enclosed by the
closed loop formed by taking instanton 1 from $\nhat_i$ to $\nhat_f$, and
instanton 2 back to $\nhat_i$. Therefore, we can write
\beq
\Dta = 2 D_1 e^{-{\rm Re\,}\Scl_1} \cos(J\Phi) + D_3 e^{-\Scl_3}
          + D_4 e^{-\Scl_4}.    \label{RDta}
\eeq
All the quantities in this equation depend on the field $H_x$. We discover that
$\Scl_3 < \Scl_4$ for all fields. At low fields, ${\rm Re}\Scl_1 < \Scl_3$, while
at high fields, the inequality is reversed.
The two actions are equal at $h_0 = 0.25$ (see \fno{Re_s1}).
\begin{figure}
\epsfig{figure=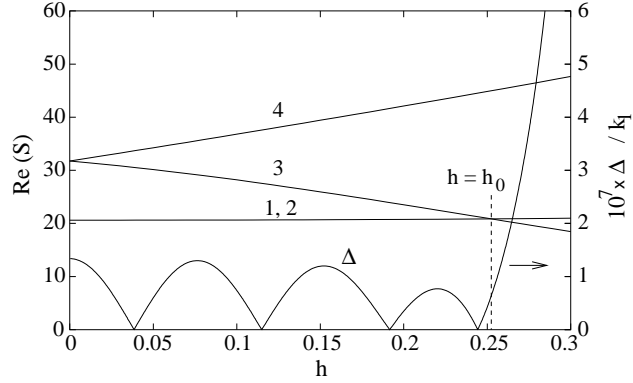}
\caption{\label{Re_s1} 
Real parts of the actions versus the magnetic field for the four instantons
in the \Fe8 problem. Also shown is the energy splitting between the ground
level pair. For $h > h_0$, instanton number 3 is the dominant one, and since
it has no interfering partner, the splitting rises with $h$ instead of
oscillating.}
\end{figure}
The prefactors
$D_i$ set the dimensional scale for the tunneling, and are generally equal to
classical small oscillation frequencies in order of magnitude. Thus we do not
expect the $D_i$ to be very different, and the relative importance of the
different instantons is determined mainly by the $S_i$. Hence, ignoring a very
small region in the immediate neighborhood of $h_0$, we can write
\beq
\Dta \apx \cases{ 2D_1 e^{-{\rm Re\,}\Scl_1} \cos(J\Phi),
                                   &  $h < h_0$;\cr
                  D_3\, e^{-\Scl_3}, & $h > h_0$. \cr}
  \label{Dtah0}
\eeq
In particular, there will be no quenchings in $\Dta$ for $h_x > h_0$.
For $h_x < h_0$, quenching will occur when $\Phi = (2n + 1)\pi/2J$, where
$n$ is an integer.

\section{Analysis for \Fe8}
\label{fe8anal}
\subsection{The interfering instantons}
\label{period}

We now turn to finding the instantons explicitly. For this purpose, it is better
to use a coordinate system having $z$ as the hard axis and $x$ as the easy axis.
Introducing spherical polar coordinates $\tta$ and $\phi$ in the standard way,
the energy (expectation value of the Hamiltonian) in this frame is given by
\bea
 E (\tta, \phi) &=&  \lam \sin^2\tta \sin^2\phi + \cos^2\tta - 2 h \cos\tta \nnu\\
                &&\quad -2 \lam_2 (\cos^4\tta + \sin^4\tta \sin^4\phi -
                                     6\sin^2\tta\cos^2\tta\sin^2\phi). 
\label{Eins2}
\eea

The first step is to find the minimum of this energy. Setting
$\ptl E/\ptl\tta$ and $\ptl E/\ptl \phi$ to zero, we obtain
\beq
\cos\phi \sin\phi \lf[ 24 \lam_2 \sin^2\tta \cos^2\tta
     -8 \lam_2 \sin^4\tta \sin^2\phi +  2 \lam \sin^2 \tta \rt] = 0
\label{eq01}
\eeq
and
\bea
\sin\tta \bigl[ 2 \lam \cos\tta \sin^2\phi &-& 2 \cos\tta +  2h  
     + 8 \lam_2 \cos^3 \tta \bigr. \nnu\\
     &-& \bigl. 8 \lam_2 (\sin^2\tta \sin^2 \phi +
                    6\sin^2\tta - 1)\cos\tta\sin^2\phi \bigr] = 0.
\label{eq02}
\eea
When these conditions are examined carefully, it is found that the minima
occur when
$\phi=0,\pi$ and the expression in the square brackets in the second equation
is zero, i.e., at $\phi=0,\pi$ and at $\tta=\tta _0$ where $\tta _0$ obeys
\beq
\cos\tta _0 - h - 4 \lam_2 \cos^3\tta _0  = 0,
\label{minim}
\eeq 
The minimum energy is
\beq
E_{\rm min} = \cos^2\tta_0 - 2 h \cos \tta_0 -2 \lam_2 \cos^4 \tta_0,
\label{minen} 
\eeq
Since $\zeta = 4 \lam_2 h^2 \ll 1$ for all $h$, one can solve \eno{minim}
perturbatively to get $\cos\tta _0 = h + 4 \lam_2 h^3$ to first order in
$\zeta$. In the same approximation, $E_{\rm min} = -h^2 - 2\lam_2 h^4$. 

Next, let us find the instantons. The trajectory, i.e., the path in phase space
without regard to the time dependence, can be found by exploiting energy 
conservation. With the abbreviations
\beq
u = \cos\tta, \quad s = \sin\phi, \label{ab_s}
\eeq
the condition $E(\tta, \phi) = E_{\rm min}$ can be written as
\beq
g(u,s) = -\tshf Z(s) u^4 + R(s) u^2 - 2h u + W(s) = 0, \label{gus}
\eeq
where
\bea
Z(s) &=& 4\lam_2 (1 + 6s^2 + s^4), \\
R(s) &=& 1 - \lam s^2 + 12 \lam_2 s^2 + 4\lam_2 s^4, \\
W(s) &=& g_0 + h^2 + \lam s^2 - 2\lam_2 s^4,
\eea
with
\beq
g_0 = -(\lam + h^2) -\emin \approx 2\lam_2 h^4 + O(h^6). \label{g0def}
\eeq

\Eno{gus} may be solved as a quartic equation for $u(s)$. Let us first
consider only real values of $\phi$, and thus of $s$.
It is useful to sketch $g(u,s)$ as a function of $u$ for real $u$,
and fixed $s$,
remembering that $|s| \le 1$ (see \fno{gus_fig}).
\begin{figure}
\epsfig{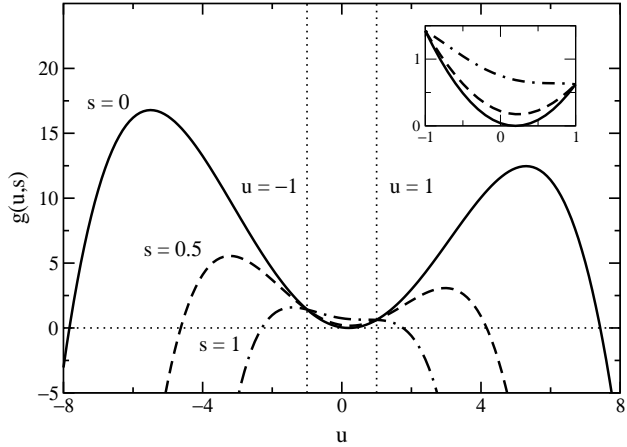}
\caption{\label{gus_fig} 
Plot of $g(u,s)$ versus $u$ for various $s$. The points to note are that
(i) except when $s=0$, $g$ has no zeros in the interval $[-1,1]$,
(ii) $g$ always has two zeros outside this interval. For $s=0$, $g$ has a double
zero at some $u \in [-1,1]$. The inset shows an enlarged view of
$g$ in the same interval.
                                 }
\end{figure}
Consider first the interval
$-1 \le u \le 1$, corresponding to points $(\tta,\phi)$ on the real unit
sphere. Since $g = E - E_{\rm min}$, it must be nonnegative in this interval.
In fact, for $s \neq 0$, it must be {\it strictly\/} positive, and for
$s=0$, it vanishes only at one point, $u = \cos\tta_0$, where it has a double
zero. Secondly, $g \to -\infty$ as $u \to \pm \infty$. It follows that
for $|s| \le 1$, $g(u,s)$ always has exactly two real roots $u(s)$, with
$|u(s)| > 1$. These roots can not be the instantons that tend to the true
energy minima as $t\to \pm\infty$. Those roots are the complex
conjugate pair with both real and imaginary parts.
 
For the complex roots let $u = A + i B$ with $A$ and $B$ being real.
From the imaginary part of \eno{gus} we obtain the equation
\beq
- 2Z(A^3 B - AB^3) + 2R AB - 2h B = 0,
\label{impart}
\eeq
while the real part yields
\beq
-\tshf Z (A^4 - 6A^2 B^2 + B^4) + R(A^2 - B^2) - 2 h A + W = 0.
\label{repart}
\eeq
Since we are not interested in solutions with either $A=0$ or $B=0$,
\eno{impart} implies that
\beq
B^2 = A^2 - (R/Z) + (h/AZ). \label{BofA}
\eeq
Substituting this into \eno{repart}, we obtain an equation for $A$ alone:
\beq
4Z^2 A^6 - 4RZ A^4 + (R^2 + 2WZ) A^2 - h^2 = 0, \label{master}
\eeq
which is a cubic in $A^2$.

When $\lam_2 = 0$, $Z(s) = 0$, and \eno{master} has the solution
$A = h/R =  h / (1-\lam \sin^2 \phi)$. We seek that solution of the cubic
which tends to this solution as $\lam_2 \to 0$. We can obtain this
approximately if we assume that $A = O(h)$. The the terms $ZA^4$ and
$Z^2 A^6$ are of order $\zeta$ and $\zeta^2$ respectively, relative to
the remaining two terms in \eno{master}. If we drop the former two terms,
the remaining equation is trivially solved to obtain
\beq
A = {h \over (R^2 + 2WZ)^{1/2}}, \label{sola1}
\eeq
which is in fact of $O(h)$. Thus, our assumption is self-consistently
verified, and the solution has the correct behavior as $\lam_2 \to 0$.

Next, we note that
\beq
R^2 + 2WZ = P'_0 + P'_2 \sin^2\phi + P'_4 \sin^4\phi,
\eeq
where we have unabbreviated $s$, and
\bea
P'_0 &=& 1 + 8 \lam_2(h^2 + g_0) \apx (1 +  \zeta)^2, \\ 
P'_2 &=& -2\lam + 24 \lam_2 + 8 \lam \lam_2 + 12 \zeta + 48\lam_2 g_0,  \\
       &\apx& -2\lam + 24\lam_2 + 8 \lam_2 \lam + 12 \zeta + 6 \zeta ^2,  \\
P'_4  &=& \lam^2 + 8 \lam_2 + 24\lam\lam_2 + 128 \lam_2^2+ 2 \zeta + 8\lam_2 g_0.\\
       &\apx& \lam^2 + 8 \lam_2 + 24 \lam \lam_2 128 + \lam_2^2+ 2 \zeta + \zeta^2.
\eea 
All three coefficients depend on $h$ only through the combination
$\zeta = 4\lam_2 h^2$, which is very small. If we neglect this weak
dependence, we get $A \apx A_0$, where
\beq
A_0 = \frac {h}{ (1 + P_2 \sin^2\phi + P_4 \sin^4\phi)^{1/2}},
\label{sola2}
\eeq
with
\bea
P_2 &=& -2\lam + 24 \lam_2 + 8 \lam_2 \lam,  \\
P_4 &=& \lam^2 + 8 \lam_2 + 24\lam\lam_2 + 128 \lam_2^2.
\eea 

For completeness, we mention that it is possible to systematically obtain
corrections to $A$ in powers of $\zeta$, in the form
$A_0 + \zeta A_1 + \cdots$. We do not carry out this exercise here.
To see how good \eno{sola2} is, we plot the real part of $\cos\tta$
in \fno{Recos}.
\begin{figure}
\epsfig{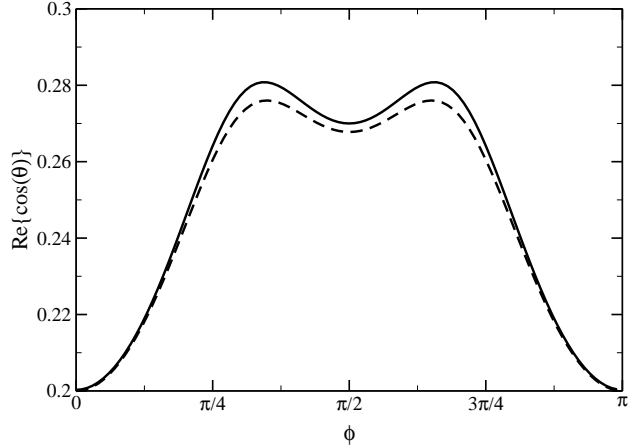}
\caption{\label{Recos} 
Plot of ${\rm Re}\,\cos\tta$ versus $\phi$ for $h=0.2$ and the \Fe8
parameters. The solid line is from exact numerical solution of $g(u,s) = 0$,
and the dashed line is the approximation (\ref{sola2}).
                                 }
\end{figure}
The dashed line is our approximation, and the
solid line is obtained by numerically solving $E-E_{\rm min}=0$ with the
numerically exact value of $E_{\rm min}$. In this plot, $h$ is taken to
be $0.2$. It can be seen that the two curves agree very well. 
   
With \eno{sola2}, we can now calculate the imaginary part of the tunneling
action, and the Berry phase $2J\Phi$ of \eno{BP}. We have
\bea
\Phi &=& \int _0^\pi (1- A)\, d\phi \nnu\\
     &\apx&   [\pi - h I(\lam,\lam_2)],
\eea
where
\beq
I(\lam,\lam_2) = \int_0^{\pi}
                   \frac {1}{ (1 + P_2 \sin^2\phi + P_4 \sin^4\phi)^{1/2}}
                        d\phi .
\label{perint}
\eeq
Various formulas for this integral are given in the Appendix.
%\ref{intappend}.

The quenching condition $J\Phi = (n +\hf)\pi$ gives the diabolical
points as
\beq
h = \frac{(2J-2n-1)\pi}{2JI},      \label{diabs}
\eeq
where $n$ is an integer. For \Fe8, $I =3.88$ as stated in
Sec.~\ref{intro}. The observed diabolical points agree extremely well
with \eno{diabs}. In particular, \eno{diabs} gives the period
$\Dta H = 0.409$~T. The experimental period is $0.41$~T.

We can now see why the diabolical points in \Fe8 are so regularly
spaced. This is because $A$ is linear in $h$ to very good approximation.
Corrections to \eno{diabs} can be found by including the correction $A_1$.
This may be important for the new system in which $\Dta$ oscillations are
indicated \cite{mn12new}.
%The ratio is $\pi / (I\sqrt{1-\lam})=1.552366261$. 

At this point, we could use \eno{BofA} and our approximation
$A \apx A_0 + \zeta A_1$ to find $B(\phi)$, and thus ${\rm Re\,}\Scl_1$.
Since the resulting analysis is not completely analytical, we forgo it,
and instead, solve for the instanton trajectory $\cos\tta(\phi)$ and
evaluate the integral for the action $S_1$ numerically.
The instantons are shown in Fig.~3 of Ref.~\onlinecite{ek+ag2}, and
the resulting approximation for $\Dta$ is shown in \fno{split_compar}.
\begin{figure}
\epsfig{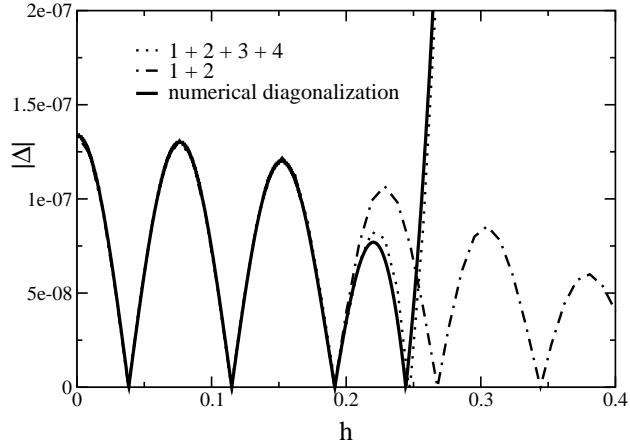}
\caption{\label{split_compar} 
Tunnel splitting (in Kelvin) between ground level pair for the model (\ref{ham1}),
as computed by numerical diagonalization of the Hamiltonian (solid line),
in the instanton approximation keeping only instantons 1 and 2 (dot-dashed
line), and in the instanton approximation with all four instantons (dotted
line). We take all prefactors $D_i$ to be equal and independent of $h$,
and adjust the common value so as to obtain the correct answer for $\Dta$
at $h=0$.
                                 }
\end{figure}
We do not know
the prefactor $D_1$, but it is clear that it is a very good approximation
to take it to be independent of $H$. It is also clear, however, that
the interfering instantons can not account for the behavior of $\Dta$ for
$h > h_0$. For that, we must turn to the boundary jump instantons.

\subsection{The Boundary-Jump (Noninterfering) Instantons}
\label{bjinsts}

The instantons we have found above have the properties (a) $\baz_i = z_i^*$,
$z_f = (\baz_f)^*$, and (b) $H(\baz_i, z_i) = H(\baz_f, z_f) = \emin$.
As discussed earlier, however, only $z_i$ and $\baz_f$ are fixed, and one
need not have $\baz_i = z^*_i$ or $z_f = (\baz_f)^*$.
Let us suppose that we start the evolution of the Euler-Lagrange equations
(\ref{ELzz}) from a point $(\baz_i,z_i)$, with $\baz_i \neq z^*_i$. Whatever
this point is, it follows from \eno{econs} that the energy $H(\baz,z)$ must
be conserved along the trajectory, and must be equal to $E = H(\baz_i,z_i)$.
Conversely, if the value of $E$ is known, we can determine the possible values
of $\baz_i$ by solving the equation $H(\baz, z_i) = E$. For the
non-boundary-jump instantons, $E = \emin$. The issue is what value of $E$
we should use for the instantons with jumps.

The answer is that we must take $E = \emin$ for the jump instantons too.
The easiest way to see this is in Klauder's formalism. His extra kinetic
term is
\beq
S'_K = 4\eps \int_0^T {1 \over (1 + \baz z)^2} {\dot \baz} \dot z dt.
   \label{SKKl}
\eeq
It is easy to write down the new Euler-Lagrange equations, and see that
energy is once again conserved. Suppose there is a boundary layer in the
solution of the EL equations around $t=t_0$. Since, as $\eps \to 0$, the
boundary layer turns into a jump, and the extra kinetic energy vanishes
for times $t = t_0+$ or $t_0-$, the energy before and after the jump must
be the same.

The problem of finding the boundary-jump instantons (indeed, all
instantons) can therefore be posed as follows. Let the ``classical"
Hamiltonian $H(\baz', z)$ have minima at $(z^*_i, z_i)$, $(z^*_f, z_f)$,
and let its value at these points be $\emin$. Then we find all possible
values of $\baz_i$ by solving the equation
\beq
H(\baz, z_i) = \emin.
\eeq
This equation has a double root at $\baz = z^*_i$, since
\beq
\lf.{\ptl \over \ptl {\bar z}} H({\bar z}, z)
                                    \rt|_{z^*_i,z_i}
 = \ \
\lf. {\ptl \over \ptl z} H({\bar z}, z)
                                    \rt|_{z^*_i,z_i}
 = \ \ 0. \label{vifeq0}
\eeq
It may, however, have additional roots at $\baz_i \neq z^*_i$. These
will be the initial points of boundary jump instantons. (Final points
$z_f$ are obtained in exactly the same way by solving
$H(\baz_f, z) = \emin$.) We then find all possible instantons $\baz(z)$
as solutions to the equation $H(\baz,z) = \emin$, and identify which
solution connects on to which end point. The time dependence is not
needed to compute the action. For, the kinetic term can be written as
\bea
S_K &=& -J\int_{z_i}^{z_f} {1 \over 1 + z \baz(z)}
                  \lf[ z {d\baz \over dz} - \baz(z) \rt] dz \\
    &=& iJ \int_{\bap_i}^{\bap_f} (1 - \cos\bat) d\bap,
\eea
the dynamical term equals
\beq
S_D = \emin T
\eeq
for all instantons, and the boundary term $S_B$ depends only on the
boundary values, vanishing for the instantons without jumps.

For our \Fe8 Hamiltonian, the argument following
Eqs.~(\ref{gus})--(\ref{g0def}) shows that we have already found all the
instantons without jumps. To find the jump instantons, let us first note
that ($\tta_i=\tta_0$, $\phi_i = 0$), and
($\tta_f = \tta_0$, $\phi_f = \pi$). The constraints on $z_i$ and $\baz_f$
therefore reduce to
\bea
\tan \tshf\bat_i e^{i\bap_i} &=& \tan \tshf\tta_0, \label{bci} \\
\tan \tshf\bat_f e^{-i\bap_f} &=& -\tan \tshf\tta_0.
\label{bcf}
\eea
Because of the symmetry of the problem, solving either of these equations will 
be enough for deducing the solution of the other. Let us solve for the
initial values. First, \eno{bci} can be solved for
$\sin\bap_i$ to yield
\beq
\sin^2\bap_i = -\lf(\frac{\cos \bat_i 
                           - \cos \tta_0}{\sin\bat_i \sin\tta_0}\rt)^2.
\label{sinp}
\eeq
Substituting this formula and \etwo{Eins2}{minen} into the energy
conservation condition $E(\bat_i, \bap_i) = \emin$ yields
\bea
0 = &-& \lam \lf(\frac{\cos\bat_i -\cos\tta_0}{\sin\tta_0}\rt)^2
   + (\cos^2\bat_i - \cos^2\tta_0) - 2 h (\cos\bat_i - \cos\tta_0) \nnu\\
    &-& 2 \lam_2 \lf[\cos\bat_i^4 +
               \lf(\frac{\cos\bat_i -\cos\tta_0}{\sin\tta_0}\rt)^4 
                 +6 \cos^2\bat_i
                     \lf(\frac{\cos\bat_i -\cos\tta_0}{\sin\tta_0}\rt)^2
              - \cos^4 \tta_0 \rt].
\label{jieqn}
\eea
We now eliminate $h$ from this equation using \eno{minim}. 
After some straightforward but lengthy algebra, we obtain
\beq
(\cos \bat_i - \cos \tta_0)^2 \lf[ 1 - {\lam \over \sin^2\tta_0}
                                      -2 \lam_2 b(\tta_i,\tta_0) \rt]
              = 0, \label{jieqn2}
\eeq
where
\beq
b(\tta_i,\tta_0) = \cos^2 \bat_i + 2 \cos \bat_i \cos \tta_0
                       + 3 \cos^2 \tta_0
                       + {(\cos \bat_i - \cos\tta_0)^2 \over \sin^4\tta_0} 
                       + 6 {\cos^2\bat_i \over \sin^2\tta_0}.
\label{defb}
\eeq
There are four solutions to \eno{jieqn2}. The first two are the non-jump
solutions, $\cos \bat_i = \cos \tta_0$. The other two, which are the
jump solutions obey
\beq
1 - {\lam \over \sin^2\tta_0} - 2 \lam_2 b(\tta_i,\tta_0) = 0.
    \label{bjieqn}
\eeq
This is a quadratic equation for $\cos\tta_i$. For $h < 1$ and small
$\lam_2$ it is easy to check that in both solutions, $\cos\bat_i$ is real
and greater than unity. Hence, we may write $\bat_i = i\nu_0$, where
$\nu_0$ is real. \Eno{bci} then shows that $\bap_i$ may be taken in the
form $\hf\pi - i\mu_0$, with $\mu_0$ real.

It is easy to see that Eqs.~(\ref{jieqn})--(\ref{bjieqn}) continue to
hold if $\tta_i$ is replaced by $\tta_f$. Thus the possible values for
$\tta_f$ are the same as those for $\tta_i$. Numbering the instantons
in question 3 and 4, we have either $\tta_f^{(3)} = \tta_i^{(4)}$, or
$\tta_f^{(3)} = \tta_i^{(3)}$. Symmetry suggests (and explicit numerics
verifies) that the latter possibility is the correct one. If we then
divide \eno{bcf} by \eno{bci}, we see that $\phi_f = \pi - \phi_i$.
Thus, the end points are related by the symmetry of reflection in the
hard-medium plane, i.e., $(J_e, J_m, J_h) \to (-J_e, J_m, J_h)$, where
the suffixes {\it e\/}, {\it m\/}, and {\it h\/} stand for easy, medium,
and hard.

If we parametrize $\bat(t)$ and $\bap(t)$ as $i \nu(t)$ and
$(\pi/2) + i \mu(t)$, then it easy to verify from the equations of motion
that $\mu(t)$ and $\nu(t)$ stay real at all $t$. With this
parametrization, the kinetic term in the action for instantons 3 and 4
may be written as
\beq
S_K = J \int_{-\mu_0}^{\mu_0} [\cosh\nu(\mu) -1 ] d\mu,
\eeq
which is real. The boundary contribution is 
\beq
S_B = 4J \ln \lf[ {\cosh(\nu_0/2) \over \cos(\tta_0/2)} \rt],
\eeq
which is also real. Hence the total action for the jump instantons is real.

The explicit calculation of the actions
must be done numerically. We solve for the trajectories in the form
$\nu(\mu)$ using energy conservation, making sure that the end points
are correct. All these calculations are done as a function of $h$. The
results for $\Scl_k$ have already been shown in \fno{Re_s1}. We can also
calculate the splitting using \eno{Dtah0}, taking $D_3 = D_1$, and
fixing $D_1$ as before. The result is shown in \fno{split_compar}.
As can be seen, the agreement with the exact diagonalization is rather
good. The $t$-dependence of $\bJ$ for instanton number 3 is shown in
Fig.~2 of Ref.~\onlinecite{ek+ag2}. The jumps are evident in this figure.

We have also carried out this exercise for $C= 1.2\times 10^{-5}\,$K.
The results are shown
in \fno{Re_s2}.
\begin{figure}
\epsfig{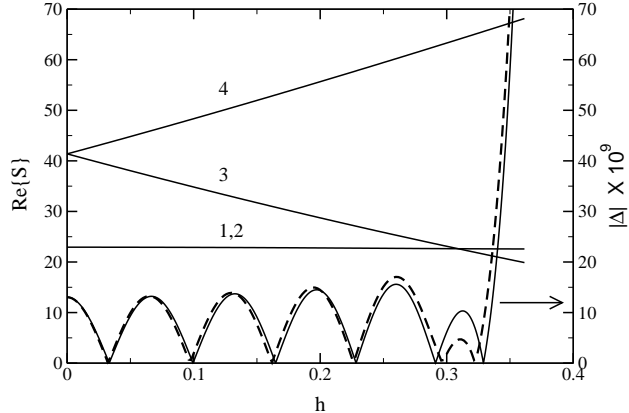}
\caption{\label{Re_s2} 
Same as \fno{Re_s1}, but for $C = 1.2 \times 10^{-5}\,$K.
In addition to the splitting $\Dta$
obtained by numerical diagonalization of the Hamiltonian matrix (solid line),
we show the answer given by \eno{RDta} (dashed line) with all prefactors
chosen to be equal, and adjusted so as to agree with the numerically
computed splitting at $h=0$.
                                 }
\end{figure}
The general quality of the instanton approach is again very
good, but it is weaker near the point where $\Scl_3 = {\rm Re}\,\Scl_1$.
The obvious reason is that we have not considered the variation with $h$
of the prefactors, especially $D_3$. If we decrease $C$ by yet another
factor of 4, \eno{RDta} with a single, $h$-independent prefactor
$D_1 = D_3$, gives six quenching points instead of eight as found
numerically. These considerations show that the prefactors are
not always unimportant; at present, however, we only know how to find
them for the non-jump instantons \cite{gkps}, and their calculation for jump
instantons is an open problem.

\subsection{Is the fourth-order anisotropy a singular perturbation?}
\label{singpert}

In the previous subsection we showed that jump instantons exist for
any $\lam_2 \ne 0$. The character of the least action or ``classical"
paths for the problem is qualitatively altered by the fourth order term in
\eno{ham1}, and from the point of view of the semiclassical analysis,
therefore, this term is a singular perturbation. Viewed as a quantum
mechanical operator, however, it is clear that the term is nonsingular;
an infinitesimal nonzero value of $\lam_2$ can not change the qualitative
nature of the energy spectrum. We therefore refer to the perturbation as
quasisingular.

It follows from this consideration that a formal analysis of the
instanton approach in the $\lam_2 \to 0$ limit will necessarily be rather
delicate. We describe briefly some analysis that shows why this is so.

We focus on jump instanton number 3, as this is clearly the most important
new contributor. For this instanton, we discover that
\beq
\mu_0 = \hf \ln \lf( {1 + u_0 \over 1 - u_0}\rt) + O(\lam_2)^{1/2},
\eeq
where $u_0 = \cos\tta_0$, which is given by \eno{minim}. Secondly, to
leading order in $\lam_2$,
\beq
\cosh \nu = {1\over \sqrt{2\lam_2}}
            \lf[ {1 - \lam\cosh^2\mu \over 1 + 6\cosh^2\mu 
                                              + \cosh^4\mu} \rt]^{1/2}.
\eeq
Because of this $\lam_2^{-1/2}$ dependence, the leading term in the
action $\Scl_3$ is also of order $\lam_2^{-1/2}$. The next term is
proportional to $J\ln\lam_2$. While the latter is well behaved as
$\lam_2 \to 0$ (remember that $\Scl$ must be exponentiated), the former
is not. It is clear that there must be a cancellation due to a corresponding
term in $D_3$, but we have not attempted to find this. This analysis shows
that the success of our assumptions about the scale and $h$-independence
of the prefactors is somewhat fortuitous. However, this assumption is
the natural one in a semiclassical approach.

\section{Number of quenching points}
\label{Nquench}

\subsection{Berry phase argument}
\label{Nq_berry}

That the tunnel splitting between two states vanishes is just another way of
saying that the states are degenerate. The magnetic fields at which such
degeneracy occurs form a set of isolated points in the magnetic field 
space ($H_x$, $H_y$, $H_z$), that are said to be {\it diabolical\/},
following Berry and Wilkinson~\cite{bwilk}. Since these points are
singularities of the energy surface, there are strong constraints on their
creation or destruction as a perturbation is continuously varied. The number
of diabolical points on the $H_x = 0$ axis is known when $C=0$; it is
interesting to inquire how many may remain when $C$ is turned on, and to
pursue this inquiry for general $J$, not just $J=10$, and also consider
fields in the {\it xz\/} plane \cite{agprev}. (Simple physical arguments
show that there can be no degeneracy if there is a component of the field
along $\yhat$.)

Let us first take $C=0$. With $H_y \ne 0$, in the standard representation
of the spin operators, the matrix of the Hamiltonian is real. By a general
theorem \cite{arnold}, the codimension of a degenerate eigenvalue
of a real Hermitean
matrix is 2. Hence, the degeneracies must occur at isolated points
in the $H_x$-$H_z$ plane. When $C$ is turned on, each one of these points
must turn into a line in the three-dimensional $(H_x, H_z, C)$ space. The
only two kinds of behaviors that are permitted by the theorem are shown in
\fno{hairpin}.
\begin{figure}
\epsfig{figure=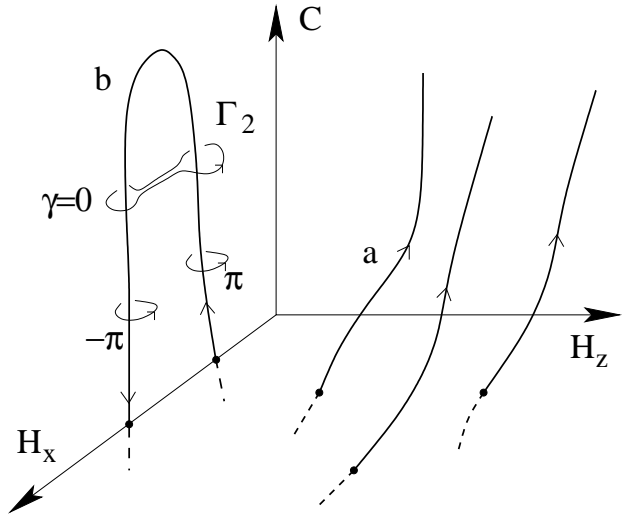,width=0.5\linewidth}
\caption{\label{hairpin} 
Trajectories of diabolical points under the influence of the fourth order
perturbation $-C(J_+^4 + J_-^4)$ in \eno{ham1}. The numbers $0$ and $\pm\pi$
are the Berry phases associated with the adjacent contours.
                                 }
\end{figure}
The first kind, marked `a', is a diabolical point that continues
on for ever. The second kind, marked `b', shows that two distinct diabolical
points in the $C=0$ plane actually lie on the same diabolical line in the
three-dimensional space. (Of course, a similar turnaround could connect the
two lines marked a and b for some $C<0$, or the line b could form a closed
loop.)

Is it possible for one of the diabolical lines to terminate abruptly?
An argument based on Berry's phase \cite{berry} shows that the answer is
no. Let $\bH$ now denote the two-dimensional vector
$(H_x, H_z)$. Let two states $\ket{\psi_1(C,\bH)}$ and
$\ket{\psi_2(C,\bH)}$ be degenerate at $\bH = \bH_0 \equiv (H_{x0}, H_{z0})$
for some value of $C$, and let $\Gam$ be a small closed contour in the
$H_x$-$H_z$ plane around the point $\bH_0$. Berry's phase, given by
\beq
\gam(\Gam) = i \oint_{\Gam} \langle \psi_1(C,\bH)
                        \ket{\nabla{_\bH}\psi_1(C,\bH)},
                \label{Bphase}
\eeq
equals $\pm\pi$ if $\Gam$ encloses a diabolical point. Otherwise,
$\gam =0$. Since a small change in $C$ or $\bH$ gives rise to a nonsingular 
perturbation of the Hamiltonian, the state $\psi_1(C,\bH)$ is a
smooth function of $C$ and $\bH$. Hence the integrand of \eno{Bphase}
can not change discontinuously under a continuous change of $C$, and the
integral must not change at all. Thus the contour $\Gam$ must continue
to encircle a degeneracy at small non zero $C$ if it did so at $C=0$.

From this point of view, the behavior `b' in \fno{hairpin} can arise only if
$\gam$ has opposite values for the two diabolical points at $C=0$. The
Berry phase for a contour $\Gam_2$ encircling both points is then 0,
and it is possible that for larger values of $C$, the contour can be
shrunk to a point without running into any singularity. Plainly, this
can happen only if the two points annihilate each other at some $C$.
We can think of simply slipping the contour $\Gam_2$ off the diabolical
line by lifting it above the hairpin bend in the figure.

It follows that diabolical points can only disappear in pairs. For the
problem of interest to us, degeneracy between ground levels, the points
are constrained to occur when $H_z = 0$ (consider
the behavior of the Hamiltonian under a $180^{\circ}$ rotation about
$\xhat$). Our analysis shows that with increasing $C$, the points at
larger $H_x$ are removed first. Since the quenching points for any
value of $J$ are located as given by \eno{DPloc} when $C=0$, it follows
that when $C\ne 0$, the number $N_q$ of such points for $H_x > 0$
must depend on $J$ as follows:
\beq
\begin{array}{ll}
2J & N_q \\
4n & {\rm\  even} \\
4n + 1 & {\rm\  even} \\
4n + 2 & {\rm\  odd} \\
4n + 3 & {\rm\  odd}
\end{array}
\label{NqvsJ}
\eeq
The same number must occur for $H_x < 0$. And, if $J$ is half-integral,
there must be a quenching point at $H_x = 0$, consistent with Kramer's
theorem. For \Fe8, $J=10$ and $N_q = 4$, which is consistent with these
general rules.

\subsection{Instanton based argument}
\label{Nq_instan}

Let us now see how our instanton analysis yields the same conclusion.
For this, let us ignore the fourth instanton, and write the amplitudes
due to the remaining instantons as \cite{Dta12}
\bea
\Dta_{1+2} &=& D_1 \bigl( e^{-\Scl_1} + e^{-\Scl_2} \bigr), \\
\Dta_3 &=&  D_3 e^{-\Scl_3}.
\eea

The keystone of our argument is the relative sign of the two
amplitudes $\Dta_{1+2}$ and $\Dta_3$, in particular, the fact that
\beq
{\rm sgn}(\Dta_{1+2}) = {\rm sgn}(\Dta_3) \ {\rm for\ large\ }h.
    \label{relsgn}
\eeq
By ``large $h$", we mean that $h$ is just less than the field strength
at which the two classical minima in the energy merge into one. We shall
not try and prove \eno{relsgn} with mathematical rigor \cite{pert}.
Rather, we argue
that it is physically plausible, for at such large fields, only spin
orientations in a very small angular range are important, and one can
use the Holstein-Primakoff, Villain, or any of a number of similar mappings
to approximate the spin operators in terms of $Q$ and $P$, position and
momentum operators for a particle in one dimension, The problem is thereby
mapped on to a particle in a double well in one dimension, for which
the splitting never vanishes. Were \eno{relsgn} not true, it is
conceivable that we could make $\Dta_{1+2} + \Dta_3$ vanish by varying
the relative value of $C$ and $(k_1 - k_2)$.

The second point is that the sign of $\Dta_3$ should not change with
$h$, since instanton 3 acts alone and has a real action. We may therefore
take $D_3 > 0$, so that $\Dta_3 > 0$ for all $h$.

The third and last point is that the sign of $\Dta_{1+2}$ at $h=0$ is now
fixed by the requirement that this amplitude vanish the correct number
of times between 0 and large $h$. Readers can verify that a correct
assignment is obtained by taking
\beq
\Dta_1 = 2 D_1 e^{-{\rm Re\,}\Scl_1} \cos(J\pi), \quad (h=0),
\eeq
with $D_1 > 0$.

The rest of the argument is simple. Continuing to ignore the fourth
instanton, the condition $\Dta = 0$ can be rewritten as
\beq
\Dta_3 = - \Dta_{1+2}. \label{zerocond}
\eeq
We now simply sketch both sides of \eno{zerocond} as a function of $h$,
keeping in mind the three points made above. This is done in \fno{nquench}
for all four classes of $J$ listed in \eno{NqvsJ}.
\begin{figure}
\epsfig{figure=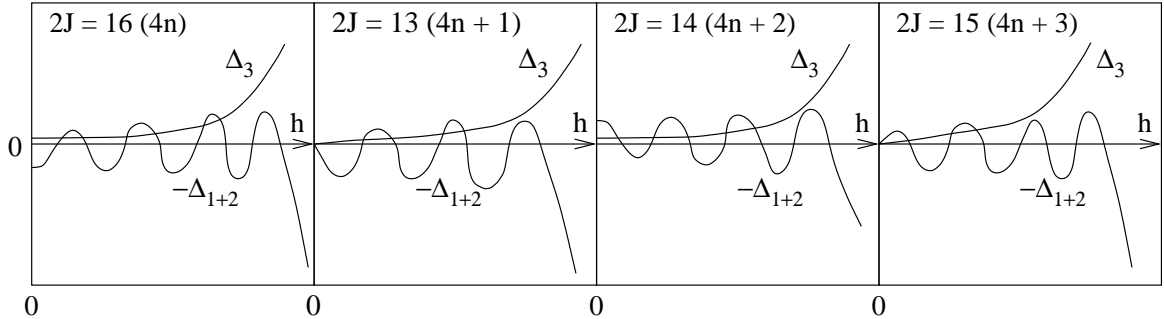,width=0.95\linewidth}
\caption{\label{nquench} 
Sketch of $-\Dta_{1+2}$ and $\Dta_3$ versus $h$ for all four classes of $J$
listed in \eno{NqvsJ}. The quenching points are given by the intersections
of these two curves. The key points to note in each case are the number of
zeros of $\Dta_{1+2}$ and the number of quenching points.
                                 }
\end{figure}
In each case, it is
obvious that the number of zeros, $N_q$, is exactly as given in this
equation. In particular, it is obvious that $N_q$ can change only in steps
of two if the curve for $\Dta_3$ is raised or lowered \cite{quadzero}.

\section{Conclusion}
\label{concl}

We have shown that the instanton formalism for spin coherent state path
integrals requires the inclusion of instantons with discontinuities at the
end points as a general matter. Such instantons are essential to understanding
the magnetic field dependence of the tunnel splitting in \Fe8. We have shown
that with certain plausible assumptions about the prefactors, the instanton
approximation can be quantitatively accurate. However, proper
calculation of tunneling prefactors for the instantons with jumps remains
an open problem.

Since jump instantons arise as a result of overspecification of boundary
conditions, and since this overspecification is a necessary consequence of
the coherent state formulation, it is clear that similar instantons will in
general be present in {\it all\/} path integrals based on coherent states.
For spin, such path integrals are unavoidable if one wishes to treat all
spin orientations on an equal footing, and they are the only way of
passing to the classical limit of the dynamics. This is not so for massive
particles. Nevertheless, there may well be some problems that are better
formulated in terms of coherent states, and then one will have to be alert
to the presence of jump instantons. An explicit instance where this is so
remains to be found.

\begin{acknowledgments}
One of us (AG) would like to acknowledge insightful discussions with
Michael Stone on spin coherent state path integrals, and Jacques Villain on
diabolical points. 
\end{acknowledgments}
\appendix*
\section{The oscillation period integral} 
\label{intappend}

In this appendix, we provide some formulas for the oscillation period
integral (\ref{perint}), which we reproduce here for convenience:
\beq
I(\lam,\lam_2) = \int_0^{\pi}
                     \frac {1}{ (1 + P_2 \sin^2\phi + P_4 \sin^4\phi)^{1/2}}
                        d\phi.
\label{perint2}
\eeq
For \Fe8, the constants $P_2$ and $P_4$ equal $-1.200$ and $0.7576$
respectively. The integrand is real with these numbers.

Let $w$ be a complex number such that
\beq
w + w^* = - P_2, \quad ww^* = P_4.
\eeq
Then, using formula 2.616.1 of Ref. \cite{gr}, we have
\beq
I =2 \int_0^{\pi} {d\phi \over [(1 - w\sin^2\phi) (1 - w^*\sin^2\phi)]^{1/2}}
  = {2 \by \sqrt{1-w}} K \lf(\sqrt{{w^* - w \by 1 - w}} \rt),
    \label{Iellip}
\eeq
where $K(k)$ is the complete elliptic integral of the first kind.

The form (\ref{perint}) shows that $I$ is real, but this is not evident
from \eno{Iellip}. But, by using the canonical form for $K$, we obtain
\beq
I = 2 \int_0^{\pi/2} {dx \by (1 - w \cos^2 x - w^* \sin^2 x)^{1/2}}.
\eeq
If we write $w = a + ib$, expand the integrand in powers of $b$, and
integrate term by term, we obtain
\bea
I &=& {\pi \by \sqrt{1-a}} \lf[ 1 - {3\by 16} \lf( b\by 1-a\rt)^2
         + \cdots + {(-1)^n \by \pi} \lf( b\by 1-a\rt)^{2n}
                  {\Gam(2n + \tshf) \by \Gam(2n + 1)}
                  {\Gam(n + \tshf) \by \Gam(n + 1)} + \cdots \rt] \nnu\\
  &=& {\pi \by \sqrt{1-a}}
         F\lf( {1\by 4}, {3\by 4}; 1; - {b^2 \by (1-a)^2} \rt),
    \label{Ihyper}
\eea
where $F$ is the hypergeometric function. This form again shows that
$I$ is real. With the \Fe8 parameters, $a = 0.5999$, and $b = 0.6307$.
The series (\ref{Ihyper}) is not convergent for these values of $a$
and $b$. For efficient numerical evaluation of $I$ in such cases, one can
either perform a numerical integration of \eno{perint} directly (which is
what we did), or reexpress the hypergeometric function in \eno{Ihyper} in
terms of other hypergeometric functions of the argument
\beq
{(1-a)^2 \by 1 - 2a + (a^2 + b^2)} = {(2 + P_2)^2 \by 4 (1 + P_2 + P_4)}
\eeq
using formula 9.132.1 of Ref. \cite{gr}. Since the argument is now small
compared to 1 (0.2870 for \Fe8), the hypergeometric series involved are
rapidly convergent.


\begin{thebibliography}{99}

\bibitem{vill} See, e.g., the review by J. Villain: {\it Molecular Magnetism:
A School of Physics\/}, in {\it Frontiers of Neutron Scattering\/}, edited by
A. Ferrer (World-Scientific, 2000).

\bibitem{werns}
W.~Wernsdorfer and R.~Sessoli, Science {\bf 284}, 133 (1999).

\bibitem{mn12new} Very recently, oscillatory tunnel splittings have been
seen in another molecule dubbed Mn$_{12}^{2-}$ by the authors:
W.~Wernsdorfer, M.~Soler, G.~Christou, and D.~N. Hendrickson, condmat/0109066.
This molecule is not yet as extensively studied as \Fe8, and the authors have
analyzed it in terms of the simplest biaxial symmetry model (\ref{ham})
given below. If further experimentation confirms fewer than the full number
of quenchings predicted by this model, that would be strong indication for
a fourth-order anisotropy term similar to that in \Fe8, and our analysis
will apply.

\bibitem{gargepl}
A.~Garg, Europhys. Lett. {\bf 22}, 205 (1993).

\bibitem{gargprl99}
A.~Garg, Phys. Rev. Lett. {\bf 83}, 4385 (1999).

\bibitem{kou}
S.~P. Kou, J.~Q. Liang, Y.~B. Zhang, and F.-C. Pu, Phys. Rev. B {\bf 59},
11792 (1999).

\bibitem{fort}
J.~Villain and A.~Fort, Euro. Phys. J. B. {\bf 17}, 69 (2000).

\bibitem{lmpp} J.-Q. Liang, H.~J. W. M\"uller-Kirsten, D.~K. Park, and
F.-C. Pu, Phys. Rev. B {\bf 61}, 8856 (2000).

\bibitem{yl} S.-K. Yoo and S.-Y. Lee, Phys. Rev. B {\bf 62}, 5713 (2000).

\bibitem{gargmathph}
A.~Garg, math-ph/0003005.

\bibitem{gargprb01}
A.~Garg, Phys. Rev. B {\bf 64}, 094413 (2001);
{\it ibid\/} {\bf 64}, 094414 (2001).

\bibitem{wilk}
M.~Wilkinson, Physica {\bf 21D}, 341 (1986).

\bibitem{ldg}
D.~Loss, D.~P. DiVincenzo, and G.~Grinstein, Phys. Rev. Lett. {\bf 69}, 3232
(1992).

\bibitem{vdh} J.~von Delft and C.~L. Henley, Phys. Rev. Lett. {\bf 69}, 3236
(1992).

\bibitem{Fadd} L.~D. Faddeev, in {\it Methods in Field Theory\/},
Les Houches 1975, edited by R.~Balian and J.~Zinn-Justin (North-Holland,
Amsterdam, 1976).

\bibitem{Klau} J.~R. Klauder, Phys. Rev. D {\bf 19}, 2349 (1979).

\bibitem{ek+ag2}
Ersin Ke\c{c}ecio\u{g}lu and A.~Garg, Phys. Rev. Lett {\bf 88}, 237205 (2002). 

\bibitem{zero_pt}
It should of course be understood that now when we speak of states in
which the spin is localized along a particular orientation, there is
always a minimum zero-point spread or uncertainty in this orientation.
Thus an equation such as $\bJ = J\zhat$ must be understood as holding
for the average $\avg{\bJ}$, and not read literally.

\bibitem{ersin}
Ersin Ke\c{c}ecio\u{g}lu and A.~Garg, Phys. Rev. B {\bf 63}, 064422 (2001). 

\bibitem{expl}
The Hamiltonian (\ref{ham1}) differs from that in Ref.~\cite{werns}
by a term proportional to $\bJ^2$. This makes no physical difference, since
$[\bJ^2, \ham] = 0$. In terms of the parameters $D$ and $E$ of
Ref.~\cite{werns}, $k_1 = D + E$ and $k_2 = D - E$.
 
\bibitem{gatt}
D.~Gatteschi, A.~Caneschi, L.~Pardi, and R.~Sessoli, Science {\bf 265}, 1054
(1994).

\bibitem{barra}
A.~L.~Barra, P.~Debrunner, D.~Gatteschi, C.~E.~Schulz, R.~Sessoli,
Europhys. Lett. {\bf 35}, 133 (1996);

\bibitem{cac}
R.~Caciuffo et al., Phys. Rev. Lett. {\bf 81}, 4744 (1998).

\bibitem{henn} M.~Hennion et al., Phys.\ Rev.\ B {\bf 56}, 8819 (1997).

\bibitem{Fel} B.~Felsager, {\it Geometry, Particles, and Fields\/}
(Springer, New York, 1998), Chapter 5.

\bibitem{hanbook} A.~Garg, Spin Tunneling in Molecular Magnets,
cond-mat/0012157.

\bibitem{fnJJ}
The state $\ket{J,J}$ is the usual simultaneous eigenstate of $J^2$ and $J_z$
with eigenvalues $J(J+1)$ and $J$, respectively. Hence, $\ket{J,J} = \ket\zhat$.

\bibitem{koch} E.~A. Kochetov, J. Math. Phys. {\bf 36}, 4667 (1995).

\bibitem{spg} M.~Stone, K.-S. Park, and A.~Garg, J. Math. Phys.
{\bf 41}, 8025 (2000).

\bibitem{fnbndry} One can also partition the action between `kinetic'
and `boundary' terms differently. Following Klauder, e.g., we could write
$S_K = -iJ\int \cos\tta d\phi$, and
$S_B = J \ln(\sin\bat_i \sin\bat_f/ \sin\tta_i \sin\tta_f)$.

\bibitem{gkps} A.~Garg, E.~Kochetov, K.-S. Park, and M.~Stone,
cond-mat/0111139.

\bibitem{bwilk}
M.~V.~Berry and M.~Wilkinson, Proc. Roy. Soc. Lond. A {\bf 392},
15 (1984).

\bibitem{agprev} The argument that follows has been partially
given previously in Ref. \onlinecite{hanbook}.

\bibitem{arnold} V.~I. Arnold, {\it Mathematical Methods of Classical
Mechanics\/} (Springer-Verlag, New York, N.Y., 1978). See Appendix 10.

\bibitem{berry} M.~V. Berry, Proc.\ R.\ Soc. Lond. A {\bf 392}, 45
(1984).

\bibitem{Dta12} It should be noted that the expression
$\Dta_{1+2} = 2D_1 \cos(J\Phi) \exp(-{\rm Re}\, \Scl_1)$ does not hold at the
fields we are considering here. As $h$ is increased starting from
0, a value $h^*$ is reached where $\Phi$ vanishes, but where the energy
still has two minima. The imaginary parts of the two instantons 1 and 2
are zero for all $h > h^*$. These points known by explicit calculation
in the case $C=0$; see A.~Garg, Phys.\ Rev.\ B {\bf 60}, 6705 (1999).

\bibitem{pert} For readers who are not happy with this, we offer the
following. In stead of fixing the relative signs of $\Dta_{1+2}$ at
large $h$, we do so at $h=0$ (or infinitesimal $h > 0$ for half-integer
$J$). This can be done by computing these amplitudes via perturbation
theory in $k_1 - k_2$, $C$, and $h$. We leave it to the readers to
do this, but similar calculations are given in 
C.-S.~Park and A.~Garg, Phys.\ Rev.\ B {\bf 65}, 064411 (2002).

\bibitem{quadzero} The count continues to be correct even in the exceptional
case in which the two curves just touch each other, if we regard this as
producing a double zero of $\Dta$. The curve of $|\Dta|$ versus $h$ will
have a quadratic zero as opposed to the cusps in Figs.~\ref{Re_s1},
\ref{split_compar}, or \ref{Re_s2}.

\bibitem{gr}
I.~S. Gradshteyn and I.~M. Ryzhik, {\it Tables of Integrals,
Series, and Products,} corrected and enlarged edition (Academic, New York,
1980).

\end{thebibliography}
\end{document}